\documentclass[12pt]{article}
\usepackage{graphicx,amsmath,amssymb}

\newcommand\pubnumber{HGU-CAP-028,SU-HET-01-2014,EPHOU-14-001}
\newcommand\pubdate{\today}

\def\shimane{Graduate School of Science and Engineering,\\
Shimane University, Shimane 690-8504, JAPAN}
\def\hokudai{Department of Physics,\\
Hokkaido University, Sapporo 060-0810, JAPAN}
\def\hokkai{Department of Life Science and Technology,\\
Hokkai-Gakuen University, Sapporo 062-8605, JAPAN}
\def\speaker{\footnote{speaker}}

\def\Title#1{\begin{center} {\Large #1 } \end{center}}
\def\Author#1{\begin{center}{ \sc #1} \end{center}}
\def\Address#1{\begin{center}{ \it #1} \end{center}}

\newcommand\pubblock{\rightline{\begin{tabular}{l} \pubnumber\\
         \pubdate  \end{tabular}}}
\newenvironment{Abstract}{\begin{quotation}  }{\end{quotation}}
\newenvironment{Presented}{\begin{quotation} \begin{center} 
             PRESENTED AT\end{center}\bigskip 
      \begin{center}\begin{large}}{\end{large}\end{center} \end{quotation}}
\def\Acknowledgements{\bigskip  \bigskip \begin{center} \begin{large}
             \bf ACKNOWLEDGEMENTS \end{large}\end{center}}

\def\beq{\begin{equation}}
\def\eeq#1{\label{#1}\end{equation}}
\def\eeqn{\end{equation}}

\def\beqa{\begin{eqnarray}}
\def\eeqa#1{\label{#1}\end{eqnarray}}
\def\eeqan{\end{eqnarray}}

\let\bar=\overbar

\def\Dslash{\not{\hbox{\kern-4pt $D$}}}
\def\dslash{\not{\hbox{\kern-2pt $\del$}}}

\def\msb{{\bar{\ssstyle M \kern -1pt S}}}

\begin{document}
\begin{titlepage}
\pubblock

\vfill
\Title{Neutrino mass from neutrinophilic Higgs and leptogenesis}
\vfill
\Author{ Naoyuki Haba}
\Address{\shimane}
\vfill
\vfill
\Author{ Osamu Seto\speaker}
\Address{\hokkai}
\vfill
\vfill
\Author{ Yuya Yamaguchi}
\Address{\hokudai}
\vfill
\begin{Abstract}
In a class of two Higgs doublet model,
 where one Higgs doublet generates masses of quarks and charged leptons 
 whereas the other Higgs doublet with a tiny vacuum expectation value(VEV) 
 generates neutrino Dirac masses,
 smallness of neutrino masses might be understand as
 the consequence of the small second Higgs VEV.
In this framework,
 thermal leptogenesis scenarios work well at low energy scale
 and have several advantages as follows. 
Under the assumption of hierarchical right-handed neutrino masses,
 the lightest right-handed neutrino can be as light as ${\cal O}(10^2)$ TeV.
The required degeneracy for successful resonant leptogenesis also
 can be significantly reduced as small as ${\cal O}(10^4)$.
Availability of low scale thermal leptogenesis
 provides a novel solution to gravitino problem
 in supergravity models.
\end{Abstract}
\vfill
\begin{Presented}
The 10th International Symposium on Cosmology and Particle Astrophysics (CosPA2013),\\
Honolulu, USA,  November 12--15, 2013
\end{Presented}
\vfill
\end{titlepage}
\def\thefootnote{\fnsymbol{footnote}}
\setcounter{footnote}{0}

\section{Introduction}

Various experiments have confirmed that neutrinos have tiny masses.
Smallness of the neutrino mass is one of the important questions
 in particle physics.
Several approaches to address this question have been proposed.
A new class of two Higgs doublet models (THDM) called
 ``neutrinophilic Higgs doublet model'', the vacuum expectation value (VEV) of
 the additional Higgs $v_{\nu}$ is of 
 much smaller energy scale comparing to the SM Higgs doublet~\cite{Ma,MaRa,Wang:2006jy,Nandi,HabaHirotsu,Marshall:2009bk},
 is also one of those candidates.
Its various aspects such as phenomenology~\cite{Davidson:2009ha,Logan:2010ag,Haba:2011nb}, vacuum structure~\cite{Haba:2011fn}, an extension to grand unification~\cite{HKS}, astrophysical and cosmological implications~\cite{Sher:2011mx,Choi:2012ap}
 have been intensively investigated.
The idea of this class of models is that 
 neutrino masses are much smaller than other fermion 
 because those come from the different Higgs doublet with a smaller VEV,
 $v_{\nu} \ll v \simeq 246$ GeV.
Masses of light neutrino are given by 
\begin{equation}
	m_{ij} = \sum_k \frac{y^{\nu}_{ik}v_{\nu} y^{\nu}{}^T_{kj}v_{\nu}}{M_k} ,
\end{equation}
 with $M_k$ being $k$-th right-handed Majorana neutrino mass.
For $v_{\nu} \ll v$,
 neutrino masses can be small for relatively large neutrino Yukawa couplings $y^{\nu}$
 and small right-handed neutrino masses $M_k$.

In modern cosmology and particle physics,
 one of important open problems
 is the origin of the baryon asymmetry in the Universe.
Thermal leptogenesis~\cite{FukugitaYanagida} 
 by heavy right-handed Majorana neutrinos for seesaw mechanism~\cite{Type1seesaw}
 is one of the most attractive scenarios for baryogenesis. 
The size of $CP$ asymmetry in a right-handed neutrino decay is, roughly speaking,
 proportional to the mass of right-handed neutrino.
For hierarchical mass spectrum of right-handed neutrino,
 the lightest right-handed neutrino
 mass must be larger than $10^8$ GeV~\cite{LowerBound,Davidson:2002qv}. 

Notice that neutrino Yukawa couplings in neutrinophilic Higgs doublet models
 do not need to be so small for lighter right-handed neutrinos. 
This fact has significant implication to leptogenesis.
This opens new possibility of low scale thermal leptogenesis. 
Here, we will show that  
 $CP$ asymmetry is enhanced and thermal leptogenesis suitably works 
 in multi-Higgs models with a neutrinophilic Higgs doublet~\cite{HabaSeto}.

The resonant leptogenesis scenario~\cite{APRD,ResonantLeptogenesis} 
 also has been known as a possibility of leptogenesis at low energy scale,
 say TeV scale, where 
 the $CP$ asymmetry is enhanced by a self energy
 of strongly degenerated right-handed neutrinos.
However, we will show 
 the masses of right-handed neutrinos can be reduced to $2$ TeV and 
 the required degeneracy becomes much milder as of order ${\cal O}(10^4)$
 in the minimal neutrinophilic Higgs doublet model~\cite{Haba:2013pca}.

The realization of low scale thermal leptogenesis is attractive
 particularly in supersymmetric models, because of ``gravitino problem''~\cite{GravitinoProblem}.
Hence, thermal leptogenesis at a low energy scale 
 in a neutrinophilic Higgs doublet model  
 offers a new solution to avoid gravitino problem~\cite{HabaSeto,HabaSeto2}.

\section{Minimal neutrinophilic THDM }
\label{sec:Minimal}

Let us show the minimal neutrinophilic THDM model originally suggested by Ma~\cite{Ma}. 
In the model, one additional Higgs doublet $\Phi_{\nu}$, which gives only neutrino Dirac masses, 
 besides the SM Higgs doublet $\Phi$ and a discrete $Z_2$-parity are introduced.
Under the discrete symmetry,  Yukawa interactions are given by 
\begin{eqnarray}
{\mathcal L}_{yukawa}=y^{u}\bar{Q}_L \Phi U_{R}
 +y^d \bar{Q}_{L}\tilde{\Phi}D_{R}+y^{l}\bar{L}\Phi E_{R} 
 +y^{\nu}\bar{L}\Phi_{\nu}N+ \frac{1}{2}M \bar{N^{c}}N
 +{\rm h.c.}\; 
\label{Yukawa:nuTHDM}
\end{eqnarray}
where 
 $\tilde{\Phi}=i\sigma_{2}\Phi^{\ast}$, and   
 we omit a generation index. 
$\Phi_\nu$ only couples with a right-handed neutrino $N$ by the $Z_2$-parity so that 
 flavor changing neutral currents (FCNCs) 
 are suppressed. 
The Higgs potential of the neutrinophilic THDM is given by 
\begin{eqnarray}
V^{\rm THDM} 
&
= m_\Phi^2 \Phi^\dag \Phi + m_{\Phi_\nu}^2 \Phi_\nu^\dag \Phi_\nu
-m_3^2\left(\Phi^\dag \Phi_\nu+\Phi_\nu^\dag \Phi\right)
+\frac{\lambda_1}2(\Phi^\dag \Phi)^2
+\frac{\lambda_2}2(\Phi_\nu^\dag \Phi_\nu)^2\nonumber \\
&\qquad+\lambda_3(\Phi^\dag \Phi)(\Phi_\nu^\dag \Phi_\nu)
+\lambda_4(\Phi^\dag \Phi_\nu)(\Phi_\nu^\dag \Phi)
+\frac{\lambda_5}2\left[(\Phi^\dag \Phi_\nu)^2
+(\Phi_\nu^\dag \Phi)^2\right]. 
\label{Eq:HiggsPot}
\end{eqnarray}
The $Z_2$-symmetry is softly broken by $m_3^2$. 

\section{Leptogenesis in neutrinophilic THDM }

\begin{figure}[t]
    \centerline{\includegraphics[width=130mm]{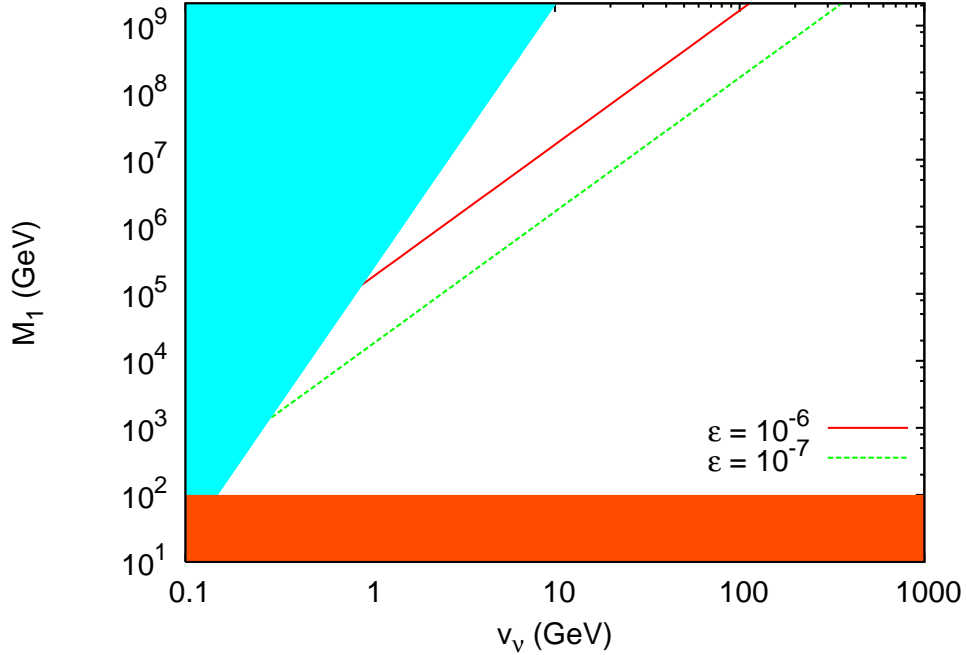}}
\caption{
 Available region for leptogenesis~\cite{HabaSeto}.
 The horizontal axis is the VEV of neutrino Higgs $v_{\nu}$ 
 and the vertical axis is the mass of the lightest right-handed neutrino $M_1$.
 In the red brown region, the lightest right-handed neutrino decay 
 is kinematically forbidden.
 In turquoise region, $\Delta L=2$ washout effect is too strong.
 The red and green line are contours of the $CP$ asymmetry
 of $\varepsilon=10^{-6}$ and $10^{-7}$, respectively.
}
\label{fig:AvailableRegion} 
\end{figure}

We consider leptogenesis in the neutrinophilic THDM with the extra Higgs doublet $\Phi_{\nu}$
 described in Sec.~\ref{sec:Minimal}.
Under hierarchical right-handed neutrino mass spectrum, 
 the $CP$ asymmetry, 
\begin{eqnarray}
\varepsilon 
 \simeq  -\frac{3}{16\pi} 10^{-6} \left(\frac{0.1 {\rm GeV}}{v_{\nu}}\right)^2
  \left(\frac{M_1}{100 {\rm GeV}}\right)
  \left(\frac{m_{\nu}}{0.05 {\rm eV}}\right) \sin\delta , 
\label{CPasym}
\end{eqnarray}
 is significantly enhanced for a light right-handed neutrinos due to
 the tiny Higgs VEV, $v_{\nu}$ .
On the other hand, one should notice that, for lower $v_{\nu}$,
 the $\Delta L =2$ lepton number violating washout processes become more significant.

All conditions for successful thermal leptogenesis is presented in Fig.~\ref{fig:AvailableRegion}.
The horizontal axis is the VEV of neutrino Higgs $v_{\nu}$ 
 and the vertical axis is the mass of the lightest right-handed
 neutrino, $M_1$.
In the red brown region, the out of equilibrium decay of lightest right-handed neutrino 
 is not possible.
In turquoise region, $\Delta L=2$ lepton number violating washout effect is too strong.
The red and green line are contours of the $CP$ asymmetry of $\varepsilon=10^{-6}$ and $10^{-7}$,
 respectively.
Thus, in the parameter region above the line of $\varepsilon = 10^{-7}$, 
 thermal leptogenesis easily works 
 even with hierarchical masses of right-handed neutrinos.

\section{Resonant leptogenesis in neutrinophilic THDM }

\begin{figure}[htbp]
 \begin{minipage}{0.48\hsize}
  \begin{center}
\includegraphics[width=70mm]{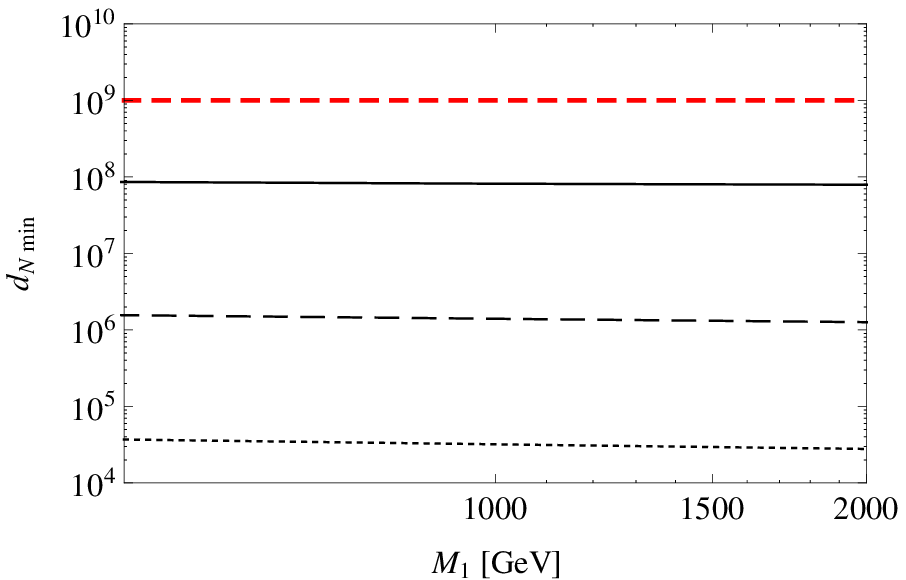}
  \end{center}
\caption{Required ${d_N}_{\rm min}$ for $K_1=10^{-2}$ and its $M_1$ dependence.
  The solid, dashed and dotted lines correspond to $y_{\nu} = 10^{-6}, 10^{-5}$,
 and $10^{-4}$, respectively. The red-thick-dashed line shows $d_N$ for
 the standard resonant leptogenesis~\cite{Haba:2013pca}. 
}
\label{fig:dmass2} 
\end{minipage}
\hfill
 \begin{minipage}{0.48\hsize} 
  \begin{center}
\includegraphics[width=70mm]{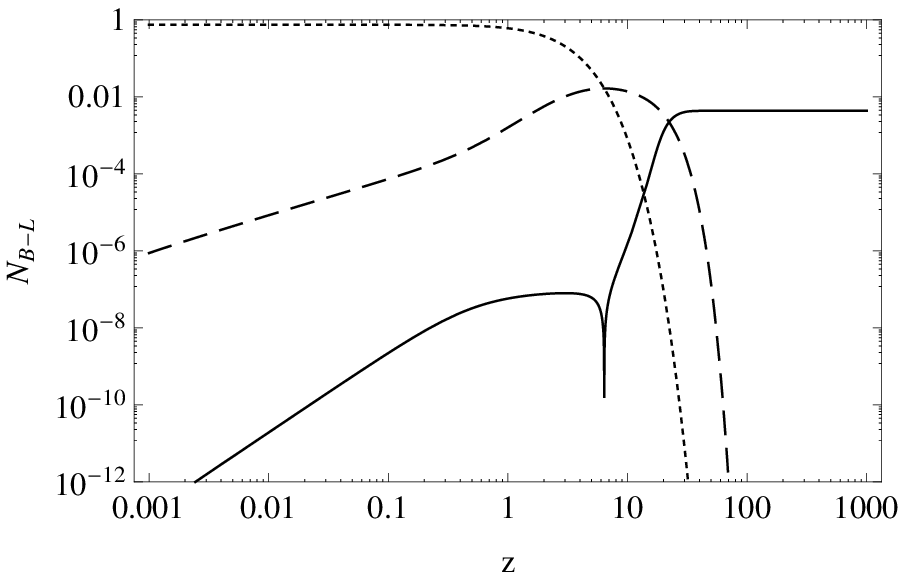}
  \end{center}
\caption{Time evolution of $B-L$ asymmetry $N_{B-L}$ with $ \varepsilon{}_1 = -1$, 
$M_1 = 2$\,TeV, $y_{\nu}=10^{-4}$, and $K_1 = 10^{-2}$ is shown by the solid line.
For information,
 the dashed and dotted lines correspond to $N_1$ and $N_2$ abundance are
 also drawn~\cite{Haba:2013pca}.
}
\label{fig:boltz} 
\end{minipage}
\end{figure}

Next, let us consider resonant leptogenesis in the neutrinophilic THDM.
When two right-handed neutrinos are degenerate, 
 the $CP$ asymmetry is approximately given by~\cite{APRD,ResonantLeptogenesis}
\begin{eqnarray}
\varepsilon_{i} &\simeq& \frac{{\rm Im}\, (y^{\nu\dagger}\,y^\nu)^2_{ij}}{
	(y^{\nu\dagger}\,y^\nu)_{ii}\,(y^{\nu\dagger}\,y^\nu)_{jj} }  
	\frac{\widetilde{m}_j M_{j}}{8\pi v^2} \frac{M_i M_j}{M_i^2 - M_j^2} ,
\label{epsN2} \\
\widetilde{m}_i &\equiv& \frac{({y^{\nu}}^{\dagger}\,y^{\nu})_{ii} v^2}{M_i },
\end{eqnarray}
 where the last factor expresses mass degeneracy of two right-handed neutrinos.
For $M_1< M_2$, here we define 
\begin{equation}
	d_N \equiv \frac{M_1 M_2}{M_2^2 - M_1^2} ,
 \label{dn} 
\end{equation}
 to parameterize the degree of degeneracy.
In neutrinophilic Higgs model,
 $v$ in Eq.~(\ref{epsN2}) is replaced with $v_{\nu} \ll v$.
Thus, from Eqs.~(\ref{epsN2}) and (\ref{dn}), one can easily find 
 a large enough $CP$ asymmetry can be obtained for smaller $d_N$.
Figure~\ref{fig:dmass2} shows that the minimum degeneracy $d_N$
 significantly reduces for larger $y^{\nu}$, equivalently smaller $v_{\nu}$.

Another consequence on resonant leptogenesis in neutrinophilic Higgs models
 is the enhancement of $\Delta L =1$ washout effect.
As mentioned above, small $v_{\nu}$ corresponds to large $y^{\nu}$.
Such large Yukawa couplings bring $N_2$ in thermal equilibrium.
This means that $N_2$ is relatively abundant when $N_1$ decays,
 as seen in Fig.~\ref{fig:boltz}.
As the results, the $\Delta L =1$ washout effect by
 the scattering between light lepton and $N_2$ tends to be large.
In order to suppress this $\Delta L =1$ washout,
 in other words to reduce relative abundance of $N_2$ to $N_1$ after $N_1$ decay,
 we need to delay $N_1$ decay time, which corresponds to a small decay parameter
\begin{equation}
	K_1 = \frac{\Gamma_{N_1}}{H(T=M_1)} ,
\label{decpar}
\end{equation}
 with $\Gamma_{N_1}$ being the decay width of $N_1$.

To generate lepton asymmetry by $N_1$ decay before the sphaleron process ceases
 at $T_{\rm sph} \simeq 100$ GeV,
 the mass of $N_1$ has to be somewhat heavier than ${\cal O}(100)$ GeV.
Fig.~\ref{fig:boltz} shows the evolution of the lepton asymmetry for $M_1 = 2$ TeV.
In this case, 
the generation of the lepton asymmetry is completed
 at around $z = M_1/T \simeq  20$
 and in time for the sphaleron transfer of
 the lepton asymmetry into the baryon asymmetry.

\section{Leptogenesis in a supersymmetric neutrinophilic model}

\begin{figure}[bhtp]
    \centerline{\includegraphics[width=130mm]{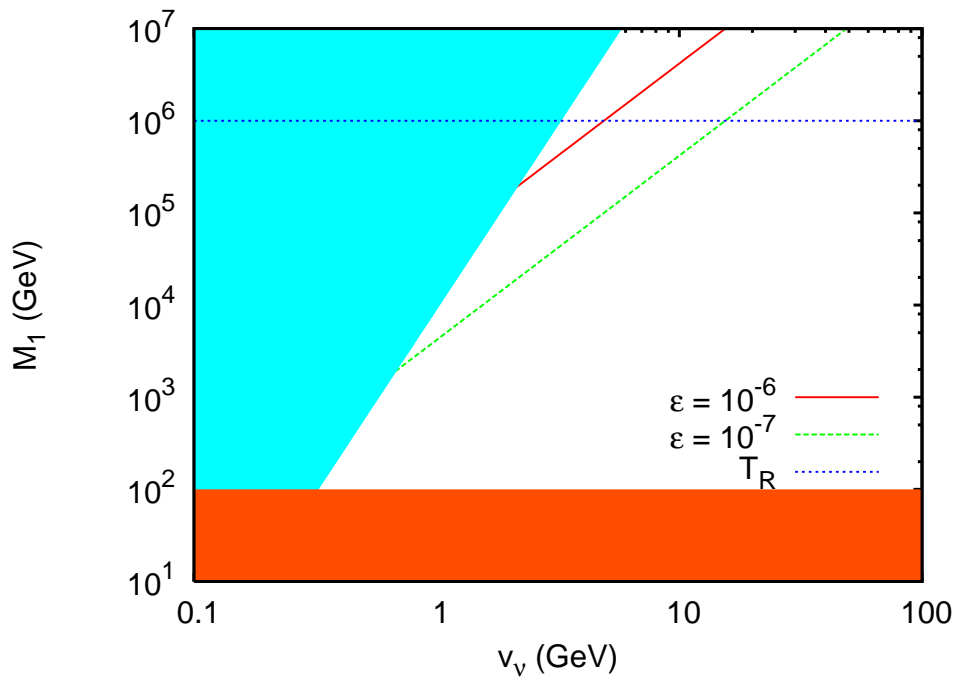}}
\caption{
 Available region for leptogenesis in a supersymmetric neutrinophilic model~\cite{HabaSeto2}.
 The blue dashed line denotes a reference value for the upper bound~\cite{GravitinoProblem2}
 of reheating temperature after inflation $T_R$.
}
\label{fig:SUSYAvailableRegion} 
\end{figure}

Construction of a supersymmetric model with $\Phi_{\nu}$ is straightforward~\cite{HabaSeto,HabaSeto2}.
By repeating the same analysis for leptogenesis in the previous section, 
 we can find the availability of thermal leptogenesis and its results 
  are summarized in Fig.~\ref{fig:SUSYAvailableRegion}.
A sufficient $CP$ violation $\varepsilon = {\cal O}(10^{-6})$ can be realized 
 for $v_{\nu} = {\cal O}(1)$ GeV 
 in the hierarchical right-handed
 neutrino with $M_1$ of ${\cal O}(10^5 - 10^6)$ GeV.
This implies
 that the reheating temperature after
 inflation $T_R$ of ${\cal O}(10^6)$ GeV is high enough 
 to produce right-handed neutrinos by thermal scatterings.
Thus, this class of model with $v_{\nu} = {\cal O}(1)$ GeV is 
 a solution to compatible with thermal leptogenesis
 in gravity mediated supersymmetry breaking with unstable gravitino.

\section{Conclusion} 

We have examined the possibility of thermal leptogenesis 
 in neutrinophilic Higgs doublet models.  
We have found 
 the available parameter region of thermal leptogenesis where 
 its washout effect is avoided as 
 keeping the $CP$ asymmetry large enough. 

We have shown the resonant leptogenesis works
 by a few TeV mass right-handed neutrino in
 the neutrinophilic THDM, with the mass degeneracy being
 of the order of only ${\cal O}(10^{4})$.

In a supersymmetric neutrinophilic Higgs doublet model,  
 we have pointed out that
 thermal leptogenesis in supergravity works well
 at low energy by a light right-handed neutrino with the mass of ${\cal O}(10^5)$ GeV.
Hence the scenario is free from gravitino problem.

\Acknowledgements
This work of N.~H. is partially supported by Scientific Grant by Ministry of
Education and Science, Nos. 00293803, 20244028, 21244036, and 23340070.


\begin{thebibliography}{99}


\bibitem{Ma}
E.~Ma, Phys. Rev. Lett. {\bf 86}, 2502 (2001). 

\bibitem{MaRa}
E.~Ma and M.~Raidal, Phys. Rev. Lett. {\bf 87} 011802 (2001), [Erratum-ibid. {\bf 87} 159901 (2001)].

\bibitem{Wang:2006jy}
F.~Wang, W.~Wang and J.~M.~Yang,
  Europhys.\ Lett.\  {\bf 76}, 388 (2006).

\bibitem{Nandi}
S.~Gabriel and S.~Nandi, Phys. Lett. B {\bf 655}, 141 (2007).

\bibitem{HabaHirotsu}
N.~Haba and M.~Hirotsu,
  Eur.\ Phys.\ J.\  C {\bf 69}, 481 (2010).

\bibitem{Marshall:2009bk}
  G.~Marshall, M.~McCaskey, and M.~Sher,
  Phys.\ Rev.\  D {\bf 81} 053006 (2010). 

\bibitem{Davidson:2009ha}
S.~M.~Davidson and H.~E.~Logan,
  Phys.\ Rev.\  D {\bf 80}, 095008 (2009). 

\bibitem{Logan:2010ag}
H.~E.~Logan and D.~MacLennan,
  Phys.\ Rev.\  D {\bf 81}, 075016 (2010). 

\bibitem{Haba:2011nb}
  N.~Haba and K.~Tsumura,
  JHEP {\bf 1106}, 068 (2011).

\bibitem{Haba:2011fn}
N.~Haba and T.~Horita,
    Phys.\ Lett.\  B {\bf 705}, 98 (2011); \\
T.~Morozumi, H.~Takata, and K. Tamai, Phys.\ Rev.\ D {\bf 85} 055002 (2012).

\bibitem{HKS}
N.~Haba,
  Europhys.\ Lett.\  {\bf 96} 21001 (2011); \\
N.~Haba, K.~Kaneta, and Y.~Shimizu,
  Phys.\ Rev.\  D {\bf 86} 015019 (2012). 

\bibitem{Sher:2011mx}
  M.~Sher and C.~Triola,
  Phys.\ Rev.\ D {\bf 83} 117702 (2011);\\
  S.~Zhou,
  Phys.\ Rev.\ D {\bf 84} 038701 (2011).

\bibitem{Choi:2012ap} 
K.~-Y.~Choi and O.~Seto,
  Phys.\ Rev.\ D {\bf 86} 043515 (2012),  [Erratum-ibid.\ D {\bf 86} 089904 (2012)];\\
K.~-Y.~Choi and O.~Seto,
  Phys.\ Rev.\ D {\bf 88}, 035005 (2013).

\bibitem{FukugitaYanagida}
M.~Fukugita and T.~Yanagida,
 Phys.\ Lett.\ B {\bf 174}, 45 (1986).

\bibitem{Type1seesaw}
P.~Minkowski,
  Phys.\ Lett.\  B {\bf 67}, 421 (1977);\\
T.~Yanagida,  
 in \textit{Proceedings of Workshop on the Unified Theory and 
 the Baryon Number in the Universe}, Tsukuba, Japan, 
 edited by A.~Sawada and A.~Sugamoto (KEK, Tsukuba, 1979), p 95; \\
M.~Gell-Mann, P.~Ramond, and R.~Slansky, 
 in \textit{Supergravity}, 
 Proceedings of Workshop, Stony Brook, New York, 1979, edited by 
 P.~Van~Nieuwenhuizen and D.~Z.~Freedman 
 (North-Holland, Amsterdam, 1979), p 315.

\bibitem{LowerBound}
W.~Buchmuller, P.~Di Bari and M.~Plumacher,
 Nucl Phys B 643, 367 (2002).

\bibitem{Davidson:2002qv}
S.~Davidson and A.~Ibarra,
  Phys.\ Lett.\  B {\bf 535}, 25 (2002).

\bibitem{HabaSeto}
  N.~Haba and O.~Seto,
  Prog.\ Theor.\ Phys.\  {\bf 125}, 1155 (2011).


\bibitem{APRD} A.~Pilaftsis, Phys.\ Rev.\ D~{\bf 56} 5431 (1997).

\bibitem{ResonantLeptogenesis}
  A.~Pilaftsis and T.~E.~J.~Underwood,
  Nucl.\ Phys.\  B {\bf 692} 303 (2004).

\bibitem{Haba:2013pca} 
  N.~Haba, O.~Seto and Y.~Yamaguchi,
  Phys.\ Rev.\ D {\bf 87}, 123540 (2013).

\bibitem{GravitinoProblem} 
M.~Y.~Khlopov and A.~D.~Linde,
  Phys.\ Lett.\  B {\bf 138}, 265 (1984) ; \\
J.~R.~Ellis, J.~E.~Kim and D.~V.~Nanopoulos,
  Phys.\ Lett.\  B {\bf 145}, 181 (1984).
 
\bibitem{HabaSeto2}
  N.~Haba and O.~Seto,
  Phys.\ Rev.\ D {\bf 84}, 103524 (2011).

\bibitem{GravitinoProblem2} 
For recent analysis, see e.g.,  \\
M.~Kawasaki, K.~Kohri, T.~Moroi and A.~Yotsuyanagi,
  Phys.\ Rev.\  D {\bf 78}, 065011 (2008) ; \\
R.~H.~Cyburt, J.~Ellis, B.~D.~Fields, F.~Luo, K.~A.~Olive and V.~C.~Spanos,
  JCAP {\bf 0910}, 021 (2009).



\end{thebibliography}
\end{document}